\documentclass[onecolumn,aps,prc,showpacs,superscriptaddress,floatfix,nofootinbib]{revtex4-2}

\usepackage{newtxtext,newtxmath,booktabs,siunitx}
\usepackage{dcolumn}
\usepackage{bm}
\usepackage{float}
\usepackage{graphicx}
\usepackage{tabularx}
\usepackage{amsmath}
\usepackage{array}
\usepackage{xcolor}
\usepackage{multirow}
\usepackage{braket}
\usepackage[colorlinks,citecolor=blue,urlcolor=blue,linkcolor=blue]{hyperref}

\begin{document}

\title{
Quantum Symmetry Restoration and Emergent Effective Deformation in Relativistic Heavy-Ion Collisions
}

\author{Hao-jie Xu}
\email{haojiexu@zjhu.edu.cn}
\affiliation{
School of Science, Huzhou Normal University, Huzhou, Zhejiang 313000, China
}

\author{Qun Wang}
\email{qunwang@ustc.edu.cn}
\affiliation{
Department of Modern Physics and Anhui Center for Fundamental Sciences in Theoretical Physics, University of Science and Technology of China, Hefei, Anhui 230026, China
}
\affiliation{School of Mechanics and Physics, Anhui University of Science and Technology, Huainan, Anhui 232001, China}

\date{\today}

\begin{abstract}
Classically deformed nuclear geometries are commonly employed in standard descriptions of relativistic collisions between two even-even nuclei, despite the fact that their exact ground states are rotationally invariant $0^+$ states.
In this paper, we formulate the collision geometry directly from the eikonal scattering matrix based on a nonorthogonal Generator Coordinate Method construction of rotationally invariant ground states.
In the optical limit, using a localized transported-density approximation for the collision-channel one-body response, rotational overlap localization generates an effective one-body density associated with the scattering process. Within this approximation, using the Gaussian Overlap Approximation and its heat-kernel representation, we show that rotational symmetry restoration acts as a geometric low-pass filter which exponentially suppresses effective deformation modes. 
The classical rigid-rotor limit is recovered for large intrinsic angular momentum fluctuations. 
We establish a microscopic framework connecting rotational symmetry restoration, collective overlap localization, and the effective deformation geometries of nuclei in high energy collisions. 
\end{abstract}

\maketitle

\section{Introduction}

The successful hydrodynamic description of anisotropic flow has 
transformed relativistic heavy-ion collisions into a sensitive probe of nuclear structure~\cite{Ollitrault1992, Heinz2013,Pang:2012he,Shen:2020mgh,Xu:2017zcn,Giacalone:2019pca, Jia:2021tzt, Ma:2022dbh,Lin:2024mnj,Xu:2024bdh,Li:2025vdp,STAR:2025elk}. 
Collective flow observables provide a mapping between initial-state spatial eccentricities and final-state momentum anisotropies, thereby offering a snapshot of the internal geometry of colliding nuclei~\cite{Giacalone:2019pca,Xu:2024bdh,Jia:2021tzt,Ma:2022dbh}. Recent STAR measurements in relativistic $^{238}$U+$^{238}$U collisions highlighted the role of nuclear deformation in determining collective flow observables~\cite{STAR:2015mki,STAR:2024wgy,STAR:2025elk}.

The standard phenomenological description of the initial state is based on Monte Carlo Glauber models~\cite{Glauber:1959,Glauber:1970jm,Miller2007,Loizides:2014vua} or energy-deposition methods such as \textsc{trento}~\cite{Moreland:2014oya}. In these models, each nucleus is represented by a classically deformed intrinsic density, typically parameterized by a Woods-Saxon profile with deformation parameters~\cite{Shou:2014eya}, and assigned a random orientation in every collision event.

This rigid-rotor picture appears conceptually inconsistent with the exact quantum state of even-even nuclei~\cite{Bohr:1951zz, Hill:1952jb, Elliott:1958zj, RingSchuck, Cline:1986ik, Agbemava:2014eja, Wang:2019dpl,DRHBcMassTable:2022uhi,Sheikh:2019qdz}. Recent discussions have emphasized that spontaneous symmetry breaking and quantum symmetry restoration are usually neglected in previous phenomenological studies of nuclear deformation in relativistic heavy-ion collisions~\cite{Duguet:2025hwi,Dobaczewski:2025rdi,Blaizot:2025bfu,Bofos:2026huw,Bofos:2026nmg}.
The physical ground state of an even-even nucleus is a rotationally invariant $0^+$ state. Consequently, the exact laboratory-frame one-body density must be spherical.

Ultrarelativistic collisions are primarily sensitive to collective geometries on the rotational manifold rather than the fully orientation-averaged laboratory-frame density.
A naive rigid-rotor treatment mixes the intrinsic broken-symmetry configuration with the exact physical state, obscuring the microscopic origin of the effective geometry sampled during the collision process.

A recent study by Ke~\cite{Ke:2025tyv} incorporated quantum orientation superposition into the Glauber model through an explicit orientation-overlap decorrelation function and showed that rotational coherence reduces the observed eccentricity in light deformed nuclei such as $^{20}$Ne. This raises the question of whether the effective geometries employed in phenomenological event generators can be systematically derived from the underlying quantum many-body theory.

In this paper, we reconstruct the initial-state geometry by evaluating the eikonal scattering amplitude on a nonorthogonal Generator Coordinate Method (GCM) rotational manifold. This approach maps the many-body scattering process onto a systematic hierarchy of effective transition densities. By employing a localized transported-density approximation together with the Gaussian Overlap Approximation (GOA) and its heat-kernel representation on the rotational manifold, we derive the effective symmetry-broken geometries that emerge in the semiclassical limit. This framework clarifies the connection between microscopic rotational coherence and macroscopic observables, establishing a theoretical foundation for nuclear geometries widely used in phenomenology within the stated approximation.

\section{Scattering matrix and effective densities on rotational manifold}

\subsection{Eikonal scattering matrix and rotational projection}

In ultrarelativistic heavy-ion collisions, the scattering process at impact parameter $\mathbf{b}$ between a projectile nucleus ($A$) and a target nucleus ($B$) is governed by the microscopic multiple scattering operator in eikonal approximation,
\begin{align}
\hat{S}(\mathbf{b})= &\prod_{i=1}^{A} \prod_{j=1}^{B} \left[ 1 - \Gamma_{\rm NN}(\mathbf{b} + \hat{\mathbf{r}}_{i\perp}^{(A)} - \hat{\mathbf{r}}_{j\perp}^{(B)}) \right], \nonumber \\ 
=& \exp \left\{ \sum_{i=1}^{A} \sum_{j=1}^{B} \ln \left[ 1 - \Gamma_{\rm NN}(\mathbf{b} + \hat{\mathbf{r}}_{i\perp}^{(A)} - \hat{\mathbf{r}}_{j\perp}^{(B)}) \right] \right\}\; ,
\label{eq:glauber_exact} 
\end{align}
where $\hat{\mathbf{r}}_{i\perp}^{(A)}$ is the transverse position operator of the $i$-th nucleon in the nucleus $A$ and $\Gamma_{\rm NN}$ is the elementary nucleon-nucleon profile function~\cite{Glauber:1959,Czyz:1969jg}. 
At the limit $\Gamma_{\rm NN}\ll 1$ we can approximate $\ln(1-\Gamma_{\rm NN}) \approx -\Gamma_{\rm NN}$ and Eq. (\ref{eq:glauber_exact}) can be put into the form
\begin{equation}
\hat{S}(\mathbf{b})\approx \exp \left\{ -\sum_{i=1}^{A} \sum_{j=1}^{B} \Gamma_{\rm NN}(\mathbf{b} + \hat{\mathbf{r}}_{i\perp}^{(A)} - \hat{\mathbf{r}}_{j\perp}^{(B)}) \right\} 
= \exp [i\hat\chi (\mathbf{b})]\;,
\label{phase}
\end{equation}
where we defined the phase-shift operator as 
\begin{equation}
\hat{\chi}(\hat{\rho}_A,\hat{\rho}_B,\mathbf{b})
= i\iint d^3\mathbf{r}_A d^3\mathbf{r}_B\,
\hat{\rho}_A(\mathbf{r}_A)\,\hat{\rho}_B(\mathbf{r}_B)\,
\Gamma_{\rm NN}(\mathbf{b}+\mathbf{r}_{A\perp}-\mathbf{r}_{B\perp})\;.
\label{eq:phase_shift}
\end{equation}
Here we introduced the microscopic one-body density operators: 
\begin{align}
\hat{\rho}_{A} (\mathbf{r}) = &\sum_{k=1}^A \delta^{(3)}(\mathbf{r} - \hat{\mathbf{r}}_k^{(A)})\;, \nonumber\\
\hat{\rho}_{B} (\mathbf{r}) = &\sum_{k=1}^B \delta^{(3)}(\mathbf{r} - \hat{\mathbf{r}}_k^{(B)})\;, 
\end{align}
where $\hat{\mathbf{r}}_k^{(i)}$ with $i=A,B$ is the $k$-th nucleon's position operator in the nucleus $i$. Throughout this work, a hat denotes an operator acting in the nuclear many-body Hilbert space, while an unhatted bold symbol denotes a spatial coordinate or a $c$-number vector. 
Then multi-body sums over nucleons in Eq. (\ref{phase}) become integrations over spatial positions of two nuclei.

The ground states of even-even nuclei are rotationally invariant: $|0_A^+\rangle$ and $|0_B^+\rangle$. In the Generator Coordinate Method (GCM)~\cite{RingSchuck}, the symmetry-restored state is
\begin{equation}
|0^+\rangle = 
\frac{1}{\sqrt{\mathcal{N}}}
\int d\mu(\Omega)\,
\hat{R}(\Omega) |\Phi \rangle,
\qquad
d\mu(\Omega)\equiv \frac{d\Omega}{8\pi^2},
\label{eq:gcm_state}
\end{equation}
where $\hat{R}(\Omega)$ is the rotation operator, $|\Phi\rangle$ is the quantum state of the nucleus in its intrinsic frame, and $\mathcal{N}$ is the normalization constant 
\begin{equation}
\mathcal{N} =
\iint d\mu(\Omega)d\mu(\Omega')\,
\bra{\Phi}
\hat{R}^\dagger(\Omega)
\hat{R}(\Omega')
\ket{\Phi}\;.
\label{eq:gcm_norm}
\end{equation}
With Eq.~\eqref{eq:gcm_state} for two colliding nuclei we obtain the exact $S$-matrix element as 
\begin{align}
\langle \hat{S}(\mathbf{b})\rangle
= & \langle 0_A^+,0_B^+ | \exp[i\hat{\chi}(\hat{\rho}_A,\hat{\rho}_B,\mathbf{b})] | 0_A^+,0_B^+ \rangle \nonumber\\
= & \frac{1}{\mathcal{N}_A\mathcal{N}_B}
\iint d\mu(\Omega_A)d\mu(\Omega'_A)
\iint d\mu(\Omega_B)d\mu(\Omega'_B) \nonumber\\
&\times
\bra{\Phi_A}\hat{R}^\dagger(\Omega_A)
\bra{\Phi_B}\hat{R}^\dagger(\Omega_B)
\exp\left[i\hat{\chi}(\hat{\rho}_A,\hat{\rho}_B,\mathbf{b})\right]
\hat{R}(\Omega'_A)\ket{\Phi_A}
\hat{R}(\Omega'_B)\ket{\Phi_B}.
\label{eq:quad_integral}
\end{align}
We see that $\langle \hat{S}(\mathbf{b})\rangle$ is the average value of $\exp[i\hat{\chi}(\hat{\rho}_A,\hat{\rho}_B,\mathbf{b})]$ weighted by intrinsic wave functions of two colliding nuclei over all orientations. 

\subsection{Effective densities}

One can expand $\hat{S}$ in Eq.~(\ref{phase}) in powers of $\hat{\chi}$ as 
\begin{equation}
\hat{S} = \sum_n \frac{1}{n!}(i\hat{\chi})^n \approx 1 + i\hat{\chi} + \cdots \;,
\end{equation}
where the $n$-th term probes $n$-body density correlations. 
For the linear term $i\hat{\chi}$,
the basic matrix element $\bra{\Phi}\hat{R}^\dagger(\Omega)\,\hat{\rho}(\mathbf{r})\,\hat{R}(\Omega')\ket{\Phi}$ is involved. 
Introducing the relative rotation $\omega \equiv \Omega ^{-1}\Omega'$, 
with $\hat{R}(\Omega') = \hat{R}(\Omega)\hat{R}(\omega)$, we obtain
\begin{align}
\bra{\Phi}\hat{R}^\dagger(\Omega)\hat{\rho}(\mathbf{r})\hat{R}(\Omega')\ket{\Phi}
= & \bra{\Phi} \hat{R}^\dagger(\Omega) \hat{\rho}(\mathbf{r}) \hat{R}(\Omega) \hat{R}(\omega) \ket{\Phi} \nonumber\ \\
= & \bra{\Phi}\hat{\rho}\left( R^{-1}(\Omega)\mathbf{r}\right) \hat{R}(\omega)\ket{\Phi}\;. 
\label{eq:matrix_raw}
\end{align}
Due to the left-invariance of the Haar measure, the integral measure can be rewritten as  
\begin{equation}
d\mu(\Omega)d\mu(\Omega') = d\mu(\Omega)d\mu(\omega).
\end{equation}
We can define the effective one-body density 
\begin{equation}
\rho_{\rm eff}(\mathbf{r})\equiv\rho_{\rm eff}^{[1]}(\mathbf{r}) 
\equiv
\frac{\int d\mu(\omega)\, \bra{\Phi}\hat{\rho}(\mathbf{r}) \hat{R}(\omega)
\ket{\Phi}}{\int d\mu(\omega)\, D(\omega)} \;, 
\label{eq:rho_mixed_def}
\end{equation}
where $D(\omega) = \bra{\Phi} \hat{R}(\omega) \ket{\Phi}$ is the overlap kernel. 
Thus we obtain 
\begin{equation}
\rho_{\rm eff}^{(\Omega)}(\mathbf{r}) \equiv \rho_{\rm eff}(R^{-1}(\Omega)\mathbf{r}) \;.
\label{rho-eff}
\end{equation}

Applying the same construction to higher-order terms in expansion of $\hat{S}$ leads to a hierarchy of $n$-body effective densities: 
\begin{equation}
\rho^{[n]}_{\rm eff}
(\mathbf{r}_1,\dots,\mathbf{r}_n)=\frac{\int d\mu(\omega)\,
\langle\Phi | \hat{\rho}(\mathbf{r}_1)\cdots\hat{\rho}(\mathbf{r}_n)\hat{R}(\omega)|\Phi \rangle}{\int d\mu(\omega)\,D(\omega)}\;.
\label{eq:n_body_eff}
\end{equation}
Standard Monte Carlo Glauber models assume independent nucleons, which is equivalent to factorizing the $n$-body density into a product of $n$ one-body densities. This retains statistical fluctuations but neglects quantum correlations present in the exact $n$-body densities.

\section{Rotational diffusion and effective deformation}

\subsection{Overlap kernel and rotational diffusion}

To evaluate the multi-dimensional orientation integral analytically and establish a practical connection with macroscopic structure parameters, we introduce a localized transported-density approximation for the one-body collision response,
\begin{equation}
\bra{\Phi}\hat{\rho}(\mathbf r)\hat R(\omega)\ket{\Phi}
\approx
D(\omega)\,
\bra{\Phi}
\hat R^\dagger(\omega)
\hat{\rho}(\mathbf r)
\hat R(\omega)
\ket{\Phi}.
\label{eq:goa_factorization}
\end{equation}
This approximation represents a finite rotational channel by the overlap kernel multiplying the rotated intrinsic density, and is not an exact identity for the microscopic off-diagonal transition density. In the localized collective regime, we subsequently use the Gaussian Overlap Approximation (GOA) to parameterize the rapidly varying collective dependence carried by $D(\omega)$. The approximation is expected to be useful in the semiclassical regime where the overlap kernel is strongly localized around small relative rotation.

To evaluate the overlap kernel analytically, we consider an axially symmetric intrinsic state that is invariant under rotations about its symmetry axis. In this case, the only independent kinematic variable is the relative polar angle $\theta$ between the respective symmetry axes. In a small-angle expansion up to quadratic order, the GOA kernel can be parameterized as~\cite{RingSchuck}:
\begin{equation}
D(\theta)
\approx
\exp\left(
-\frac{1}{2}\langle \hat{J}_y^2\rangle\theta^2
\right)\;.
\label{eq:goa_axial}
\end{equation}
Here, $\langle \hat{J}_y^2\rangle$ denotes the intrinsic fluctuation of the angular momentum in the transverse direction, which governs the angular stiffness of the quantum overlap.

To obtain an analytical solution to $\rho_{\rm eff}(\mathbf{r})$, we map the localized Gaussian profile to the short-time heat kernel $K_\tau(\theta)$ defined in the orientation space. Using the axial symmetry of the system, the isotropic diffusion equation and its singular initial boundary condition can be formulated entirely in terms of the relative polar angle $\theta$,  
\begin{equation}
\frac{\partial K_\tau(\theta)}{\partial\tau}
=
\frac{1}{\sin\theta} \frac{\partial}{\partial \theta} \left( \sin\theta \frac{\partial K_\tau(\theta)}{\partial \theta} \right)\;,
\qquad
\lim_{\tau\rightarrow0}K_\tau(\theta)=\frac{\delta(\theta)}{2\pi\sin\theta}\;,
\label{eq:heat_s2_theta}
\end{equation}
where the differential operator on the right-hand side is the angular part of $\nabla^2$. The exact solution can be expressed by an expansion in the Legendre polynomials, 
\begin{equation}
K_\tau(\theta)
=
\sum_{l=0}^{\infty}
\frac{2l+1}{4\pi}
e^{-l(l+1)\tau} P_l(\cos\theta) \;.
\label{eq:heat_kernel_s2}
\end{equation}
In the small-angle limit $\theta \to 0$, this heat kernel is reduced to the localized Gaussian form, 
\begin{equation}
K_\tau(\theta)
\propto
\exp\left(
-\frac{\theta^2}{4\tau} \right)\;.
\label{k-tau}
\end{equation}
Mapping the small-angle expression in Eq.~\eqref{k-tau} to the microscopic GOA kernel in Eq.~\eqref{eq:goa_axial} gives 
\begin{equation}
\tau = \frac{1}{2\langle\hat{J}_y^2\rangle}\;.
\label{eq:tau_sigma}
\end{equation}
The localization width of the quantum overlap kernel is then naturally connected with a rotational diffusion parameter in collective space.

The heat-kernel representation provides a compact global implementation of the local GOA smearing on the axial collective manifold.

\subsection{Effective deformation}

By substituting the expansion of the heat kernel in Legendre polynomials from Eq.~\eqref{eq:heat_kernel_s2} back into the localized GOA convolution in Eq.~\eqref{eq:rho_mixed_def} with Eq.~\eqref{eq:goa_factorization}, the effective density can be expressed as
\begin{equation}
\rho_{\rm eff}(\mathbf{r}) =
\sum_{l=0}^{\infty} (2l+1) e^{-l(l+1)\tau} \int_{0}^{2\pi} \frac{d\phi}{2\pi} \int_0^\pi \sin\theta\,d\theta\, P_l(\cos\theta) \, \bra{\Phi}\hat{R}^{\dagger}(\theta, \phi) \hat{\rho}(\mathbf{r}) \hat{R}(\theta, \phi) \ket{\Phi}.
\label{eq:rho_eff_direct_int}
\end{equation}
To evaluate this integral, we expand the intrinsic density in spherical harmonics in its body-fixed frame, 
\begin{equation}
\rho_{\rm intr}(\mathbf{r}) \equiv
\bra{\Phi}\hat{\rho}(\mathbf{r})\ket{\Phi} = \rho_0(r) +
\sum_{l} \beta_{l}^{\rm intr} \rho_{l}(r) Y_{l0}(\hat{\bf r})\;,
\label{eq:rho_intr_ylm}
\end{equation}
where $\hat{\bf r}\equiv \mathbf{r}/r$ (with $r=|\mathbf{r}|$) is a unit vector instead of an operator, and $\beta_{l}^{\rm intr}$ are expansion coefficients. 

Under a general rotation, spherical harmonics transform through the Wigner $D$-matrices, 
\begin{equation} 
Y_{l0}(R^{-1}(\omega)\hat{\bf r}) = \sum_m D^{(l)}_{m0}(R(\omega )) Y_{lm}(\hat{\bf r})\;.
\end{equation} 
By replacing a general rotation $\omega$ with a rotation $\theta$ around the $y$-axis, the rotated intrinsic density takes the form, 
\begin{align}
\rho_{{\rm intr}}(R^{-1}(\theta)\mathbf{r})\equiv & \left\langle \Phi\right|\hat{\rho}(R^{-1}(\theta)\mathbf{r})\left|\Phi\right\rangle \nonumber \\
= & \left\langle \Phi\right|\hat{R}^{\dagger}(\theta)\hat{\rho}(\mathbf{r})\hat{R}(\theta)\left|\Phi\right\rangle \nonumber \\
= & \rho_{0}(r)+\sum_{l}\beta_{l}^{{\rm intr}}\rho_{l}(r)Y_{l0}(R^{-1}(\theta)\hat{{\bf r}})\nonumber \\
= & \rho_{0}(r)+\sum_{l,m}\beta_{l}^{{\rm intr}}\rho_{l}(r)\sqrt{\frac{4\pi}{2l+1}}Y_{lm}^{*}(\theta,0)Y_{lm}(\hat{{\bf r}})\;,\label{eq:rotated_density_projection}
\end{align}
where we used $D_{m0}^{(l)}(R(\theta))=\sqrt{4\pi/(2l+1)}Y_{lm}^{*}(\theta,0)$. 
Substituting Eq.~(\ref{eq:rotated_density_projection}) into Eq.~(\ref{eq:rho_eff_direct_int}) 
and performing the angular integration, we obtain the effective density,
\begin{equation}
\rho_{\rm eff}(\mathbf{r})
=
\rho_0(r)
+
\sum_{l}
\beta_{l}^{\rm eff}
\rho_l(r)
Y_{l0}(\hat{\bf r}) \;,
\end{equation}
where effective deformation parameters emerge as  
\begin{equation}
\beta_{l}^{\rm eff}
=
\beta_{l}^{\rm intr}
\exp\left[ -\frac{l(l+1)}{2\langle\hat{J}_y^2\rangle} \right] \;.
\label{eq:beta_axial_final}
\end{equation}
We see that only the component $m=0$ contributes after integration over $\theta$ in Eq.~(\ref{eq:rho_eff_direct_int}) with Eq.~(\ref{eq:rotated_density_projection}). Equation~\eqref{eq:beta_axial_final} is obtained within the localized transported-density approximation in Eq.~\eqref{eq:goa_factorization}.

\begin{figure}[htbp]
\centering
\includegraphics[width=0.75\columnwidth]{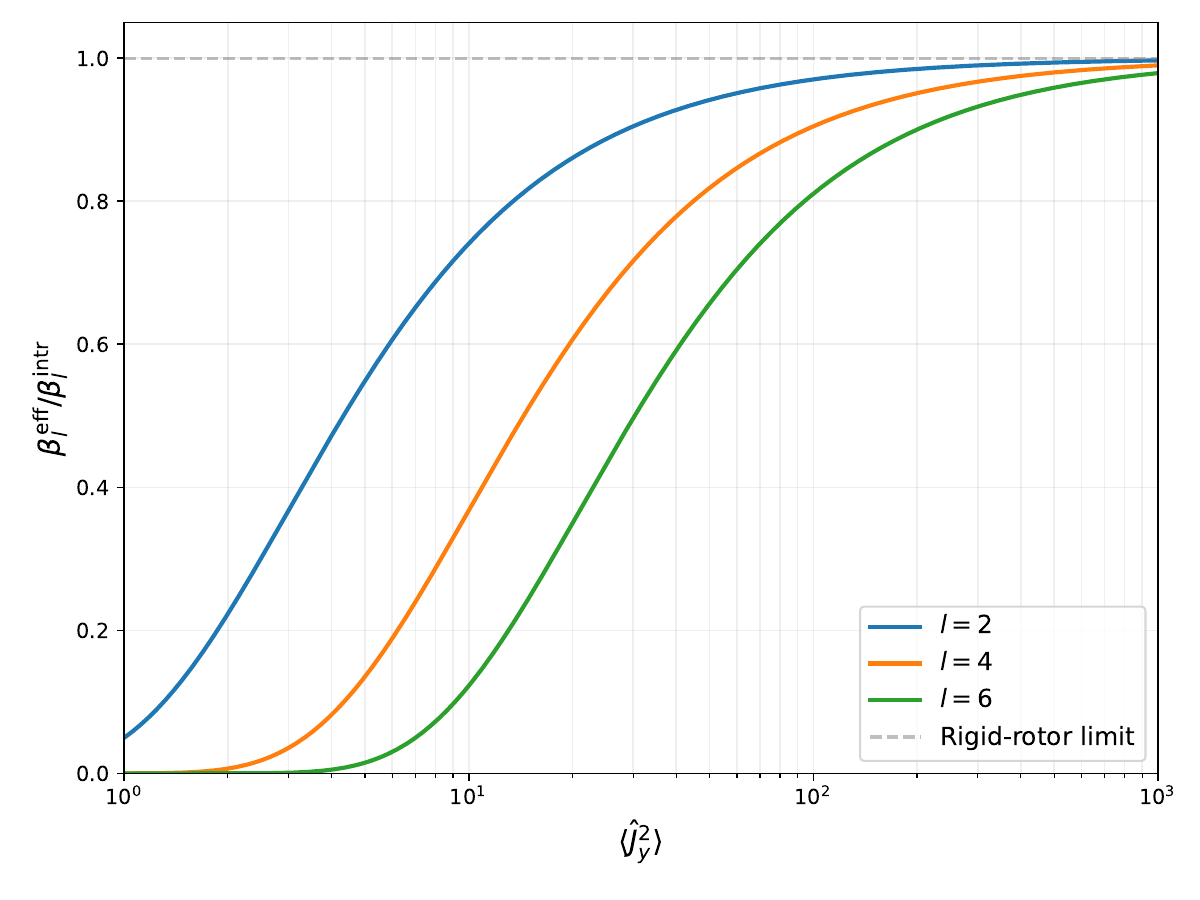}
\caption{Suppression factor $\beta_{l}^{\rm eff}/\beta_{l}^{\rm intr}$ as functions of the intrinsic angular momentum fluctuation $\langle\hat{J}_y^2\rangle$ for multipole components of $l=2,4,6$. The ratio approaches unity for large $\langle\hat{J}_y^2\rangle$, corresponding to the classical rigid-rotor limit in which finite-width overlap effects become negligible.}
\label{fig:multipole_suppression}
\end{figure}

As shown in Fig.~\ref{fig:multipole_suppression}, the overlap kernel acts as a geometric filter: higher multipoles are exponentially suppressed by rotational diffusion. The suppression is governed by $\langle \hat{J}_y^2\rangle$ rather than the intrinsic deformation alone. For highly deformed nuclei such as $^{238}$U~\cite{Ke:2025tyv} with $\langle\hat{J}_y^2\rangle\gg 1$, the classical rigid-rotor limit $\beta_{l}^{\rm eff}\rightarrow\beta_{l}^{\rm intr}$ can be reached. 

\subsection{Analytical test of geometric overlap kernel}

To test the accuracy of the GOA analytically, we utilize a functional form for the overlap kernel motivated by the algebraic structure of Elliott's SU(3) model~\cite{Elliott:1958zj}. Specifically, we consider a geometric overlap function defined on the continuous rotational manifold, 
\begin{equation}
D_{\rm exact}(\theta) = \cos^\lambda\theta \;,
\label{eq:su3kernel_elliott}
\end{equation}
whose power-law form mimics the localization structure of highest-weight states in the $(\lambda,0)$ representation. In the present work, however, this form is treated as a continuous geometric overlap kernel on the rotational manifold rather than as an exact microscopic shell-model overlap. 
For the corresponding collective description, all excitation quanta are aligned along the $z$-axis ($n_z=\lambda, n_x=n_y=0$), leading to a transverse angular momentum fluctuation of $\langle \hat{J}_y^2 \rangle = \lambda$. The naturally constructed GOA kernel is then given by
\begin{equation}
D_{\rm GOA}(\theta)
=
\exp\left(
-\frac{\lambda}{2}\theta^2
\right)\;.
\label{eq:su3goa_elliott}
\end{equation}
This matches the small-angle Taylor expansion of $\ln D_{\rm exact}(\theta)$ up to the quadratic order.

The suppression factor $R_l \equiv \beta_{l}^{\rm eff}/ \beta_{l}^{\rm intr}$ with the geometric overlap kernel in Eq.~(\ref{eq:su3kernel_elliott}) can be evaluated analytically via the Peierls--Yoccoz integral over the half-sphere $[0,\pi/2]$ as 
\begin{equation}
R_l^{\rm exact}
=
\frac{
\int_0^{\pi/2} d\theta\,
\sin\theta\,
\cos^\lambda\theta\,
P_l(\cos\theta)
}{
\int_0^{\pi/2} d\theta\,
\sin\theta\,
\cos^\lambda\theta
}
=
\frac{\lambda!!(\lambda+1)!!
}{
(\lambda-l)!!(\lambda+l+1)!!
} \;.
\label{eq:rexact_elliott}
\end{equation}
Inserting the Gaussian overlap kernel in Eq.~\eqref{eq:su3goa_elliott} directly into the Peierls--Yoccoz integral, the corresponding GOA suppression factor becomes 
\begin{equation}
R_l^{\rm GOA}
=
\frac{
\int_0^{\pi/2} d\theta\,
\sin\theta\,
\exp(-\lambda\theta^2/2)\,
P_l(\cos\theta)
}{\int_0^{\pi/2} d\theta\,
\sin\theta\,
\exp(-\lambda\theta^2/2)}\;.
\label{eq:rgoa_integral_elliott}
\end{equation}
This one-dimensional quadrature is the GOA counterpart of Eq.~(\ref{eq:rexact_elliott}). It has no equally transparent analytical expression. 
The heat-kernel representation of the GOA gives the compact expression
\begin{equation}
R_l^{\rm HK} = \exp\left[ -\frac{l(l+1)}{2\lambda} \right]\;.
\label{eq:rhk_elliott}
\end{equation}
The numerical results for $R_l^{\rm exact}$, $R_l^{\rm GOA}$, and $R_l^{\rm HK}$ in Eqs.~\eqref{eq:rexact_elliott}--\eqref{eq:rhk_elliott} are listed in Tab.~\ref{tab:su3_goa_hk} for comparison.

\begin{table}[H]
\centering
\begin{tabular*}{0.4\textwidth}{@{\extracolsep{\fill}}cccc}
\toprule
$l$ & $R_l^{\rm exact}$ & $R_l^{\rm GOA}$ & $R_l^{\rm HK}$ \\
\midrule
2 & 0.727 & 0.693 & 0.687 \\
4 & 0.336 & 0.294 & 0.287 \\
6 & 0.090 & 0.077 & 0.072 \\
\bottomrule
\end{tabular*}
\caption{Comparison of numerical results for $R_l^{\rm exact}$, $R_l^{\rm GOA}$, and $R_l^{\rm HK}$ in Eqs.~\eqref{eq:rexact_elliott}--\eqref{eq:rhk_elliott} with $D(\theta)=\cos^{\lambda=8}\theta$. The choice $\lambda=8$ is motivated by the leading intrinsic configuration in Elliott's SU(3) model commonly associated with $^{20}$Ne, while the kernel is used only for a test.}
\label{tab:su3_goa_hk}
\end{table}

Here $R_l^{\rm exact}$ refers to the result obtained with the chosen geometric overlap kernel in Eq.~\eqref{eq:su3kernel_elliott} within the transported-density approximation, rather than to an exact microscopic scattering result. As shown in Tab.~\ref{tab:su3_goa_hk}, the numerical results for GOA and heat-kernel are very close in this example, showing the exponential suppression effect in GOA. The deviation from the exact overlap kernel originates primarily from higher-order angular corrections, beginning with the $\mathcal{O}(\theta^4)$ term, which encode the non-Gaussian nature of the angular momentum distribution beyond the Gaussian and heat-kernel approximation. The present geometric overlap kernel can describe the dominant rotational localization effect for nuclear deformation with axial symmetry in the intrinsic frame.

We note that the present geometric overlap kernel is formulated for even-multipole components. However, our current approach can also be generalized to reflection-asymmetric shapes such as octupole deformation, which we leave for a future study.

\section{Incorporation of quantum correlation into Glauber model}
The effective densities derived in this work establish a bridge between the microscopic GCM-based scattering formalism and phenomenological Glauber model description of relativistic heavy-ion collisions. 

Introducing the relative rotation $\omega_A$ and $\omega_B$, the S-matrix element in 
Eq.~(\ref{eq:quad_integral}) can be put into the form
\begin{align}
\langle\hat{S}(\mathbf{b})\rangle= &\frac{1}{\mathcal{N}_{A}\mathcal{N}_{B}}\iint d\mu(\Omega_{A})d\mu(\Omega_{B})\iint d\mu(\omega_{A})d\mu(\omega_{B})\nonumber\\
	&\times\left\langle \Phi_{A},\Phi_{B}\right|\hat{R}^{\dagger}(\Omega_{A})\hat{R}^{\dagger}(\Omega_{B})\exp\left[i\hat{\chi}(\hat{\rho}_{A},\hat{\rho}_{B},\mathbf{b})\right]\hat{R}(\Omega_{A})\hat{R}(\Omega_{B})\hat{R}(\omega_{A})\hat{R}(\omega_{B})\left|\Phi_{A},\Phi_{B}\right\rangle \;.
\label{sb-rel}
\end{align}
In the optical approximation, fluctuations of the eikonal exponent are neglected within a fixed rotational channel by retaining only its first cumulant,
\begin{equation}
\left\langle e^{i\hat{\chi}}\right\rangle_{\Omega_A,\Omega_B}
\approx
\exp\left[i\left\langle\hat{\chi}\right\rangle_{\Omega_A,\Omega_B}\right].
\label{eq:optical_cumulant}
\end{equation}
Using this one-body optical truncation together with the localized transported-density approximation in Eq.~\eqref{eq:goa_factorization}, Eq.~(\ref{sb-rel}) can be approximated as 
\begin{equation}
\langle\hat{S}(\mathbf{b})\rangle_{\mathrm{opt}}\approx	
\iint d\mu(\Omega_{A})d\mu(\Omega_{B})\exp\left[i\chi_{\mathrm{eff}}\left(\Omega_{A},\Omega_{B},\mathbf{b}\right)\right] \;,
\label{sb-opt}
\end{equation}
where the effective phase-shift is defined as
\begin{equation}
\chi_{\mathrm{eff}}\left(\Omega_{A},\Omega_{B},\mathbf{b}\right)=i\iint d^3\mathbf{r}_{A}d^3\mathbf{r}_{B}\,\rho_{A,\mathrm{eff}}^{(\Omega_{A})}(\mathbf{r}_{A})\,\rho_{B,\mathrm{eff}}^{(\Omega_{B})}(\mathbf{r}_{B})\,\Gamma_{{\rm NN}}(\mathbf{b}+\mathbf{r}_{A\perp}-\mathbf{r}_{B\perp})\;.
\end{equation}
Here $\rho_{A,\mathrm{eff}}^{(\Omega_{A})}$ and $\rho_{B,\mathrm{eff}}^{(\Omega_{B})}$
are given in Eq.~(\ref{rho-eff}).  

The expression in Eq.~(\ref{sb-opt}) reproduces the orientation integration performed in Glauber-type description, with the rigid intrinsic density replaced by symmetry-restored densities. The quantity $\rho_{\rm eff}^{(\Omega)}$ is not the fully orientation-averaged laboratory density of the $0^+$ state, but a localized collective geometry governing the optical response in a fixed rotational channel in collision.

Beyond the optical approximation, the formalism generates a hierarchy of correlated effective densities $\rho_{\rm eff}^{[n]}$ in Eq.~(\ref{eq:n_body_eff}) from higher order terms of $(\hat\chi)^n$. The one-body density $\rho_{\rm eff}^{[1]}$ in Eq.~(\ref{eq:rho_mixed_def}) controls the effective deformation via Eq.~\eqref{eq:beta_axial_final}, while higher-order terms encode rotational correlations absent in the independent-particle sampling.

This distinction becomes important for event-by-event Monte Carlo implementation of the Glauber model. In the optical approximation, the hierarchy is truncated because only the mean opacity is retained. In contrast, an independent-particle Monte Carlo Glauber model uses a factorization ansatz to generate fluctuating many-body configurations. However, the exact projected hierarchy violates such a factorization 
\begin{equation}
\rho_{\rm eff}^{[n]}(\mathbf{r}_1,\ldots,\mathbf{r}_n) \neq \prod_i \rho_{\rm eff}^{[1]}(\mathbf{r}_i)\;.
\label{eq:nonfactorization_hierarchy}
\end{equation}
Therefore, replacing $n$-body effective density by a product of one-body densities neglects quantum correlation encoded in the symmetry-restored many-body state.

This reflects a group-theoretic structure of angular-momentum projection rather than ordinary dynamical nucleon correlation. Independent-particle sampling is based on $\rho_{\rm eff}^{[1]}$ and preserves classical statistical fluctuation, but it does not contain quantum correlations and fluctuations encoded in the symmetry-restored many-body state. A consistent microscopic implementation beyond the optical limit therefore requires next-generation event generators capable of sampling directly from the correlated hierarchy $\rho_{\rm eff}^{[n]}$, providing a route toward a quantum many-body treatment of the initial state in collision.

\section{Summary and Discussion}

In this work, we reformulated the initial-state geometry of relativistic heavy-ion collisions by evaluating the eikonal scattering matrix on a nonorthogonal GCM rotational manifold.
This formalism clarifies the distinction between the exact rotationally invariant laboratory-frame density and the effective geometry relevant to localized collisions in the optical limit.
Within the localized transported-density approximation, semiclassical localization of the overlap kernel generates an effective one-body density for the scattering process.

Using the Gaussian Overlap Approximation together with its heat-kernel representation on the rotational manifold, we obtained
\(
\beta_{l}^{\rm eff}
=
\beta_{l}^{\rm intr}
\exp\left[
-\frac{l(l+1)}
{2\langle\hat{J}_y^2\rangle}
\right]
\).
This relation shows that rotational symmetry restoration acts as a geometric low-pass filter on collective deformation modes.
In the large-$\langle\hat{J}_y^2\rangle$ limit, the suppression becomes weak and the classical rigid-rotor picture is recovered.
This limit corresponds to the geometry commonly employed in event generators based on phenomenological Glauber models.

The framework further leads naturally to a hierarchy of $n$-body effective densities $\rho^{[n]}_{\rm eff}$ associated with the collective rotational manifold.
Extending event generators beyond independent-particle sampling toward correlated densities may provide a systematic route to incorporating quantum many-body correlations into phenomenological models of heavy-ion collisions.

Several extensions of the present framework are possible.
Transitional nuclei and soft collective systems will require the inclusion of shape fluctuations and triaxiality~\cite{Zhao:2024lpc}.
Pairing correlations may be incorporated through Hartree--Fock--Bogoliubov intrinsic states and effective pairing tensors~\cite{Baranger:1961,Bender:2003,Zhang:2026lsg}.
The collective-kernel formalism developed in this work may also be useful for probing symmetry-restored quantum many-body systems in other collision processes, such as ultraperipheral collisions or electron-ion scattering~\cite{Wang:2021kxm,Mantysaari:2023prg,Mantysaari:2024xmy,STAR:2022wfe}.

In conclusion, we have established a microscopic framework connecting rotational symmetry restoration, collective overlap localization, and effective nuclear deformation geometries in relativistic heavy-ion collisions.

\section*{Acknowledgments}

H.X. is supported in part by the National Natural Science Foundation of China (NSFC) under Grant No. 12275082. 
Q.W. is supported in part by the National Natural Science Foundation of China (NSFC) under Grant No. 12135011. 

\bibliography{refs}

\end{document}